\documentclass[letterpaper,twocolumn,10pt]{article}
\usepackage{usenix2019_v3}

\usepackage{tikz}
\usepackage{enumitem}
\usepackage{amsmath}
\usepackage{xspace}
\usepackage{amsthm}
\usepackage{thmtools}
\usepackage{listings}
\usepackage{caption}
\usepackage{amsmath}
\usepackage{pifont}
\usepackage{subcaption}
\usepackage{sidecap}
\usepackage{wrapfig}
\usepackage{graphicx}
\usepackage{amssymb}
\usepackage{xcolor}
\usepackage{enumitem,kantlipsum}
\usepackage{wrapfig}
\usepackage{orcidlink}
\usepackage{comment}
\usepackage{xurl}
\usepackage{hyperref}

\usepackage{filecontents}

\newcommand{\ufa}{\textsc{UFA}\xspace}

\newcommand{\alwayson}{\texttt{Always-On}\xspace}%
\newcommand{\activemigrate}{\texttt{Active-Migrate}\xspace}%
\newcommand{\restorelater}{\texttt{Restore-Later}\xspace}%
\newcommand{\terminate}{\texttt{Terminate}\xspace}%

\lstdefinelanguage{CustomGo}{%
aboveskip=5pt,
belowskip=0pt,
lineskip= {-1.5pt},
language=go,                %
basicstyle=\scriptsize,       %
numbers=left,                   %
numberstyle=\tiny,      %
stepnumber=1,                   %
numbersep=2pt,                  %
backgroundcolor=\color{white},  %
showspaces=false,               %
stringstyle=\scriptsize,
identifierstyle=\scriptsize,
commentstyle=\scriptsize,
basicstyle=\scriptsize\ttfamily,
showstringspaces=false,         %
showtabs=false,                 %
frame=tb,                   %
tabsize=2,                      %
captionpos=b,                   %
breaklines=true,                %
breakatwhitespace=false,        %
title=\lstname,                 %
  sensitive,%
  morecomment=[s]{/*}{*/},%
  morecomment=[l]//,%
  morestring=[b]',%
  morestring=[b]",%
  morestring=[s]{`}{`},%
  morekeywords=[1]{break,case,const,continue,default,defer,%
      else,fallthrough,false,for,func,go,goto,if,import,iota,skip,%
      range,return,select,switch,true,type,nop,%
      var,then,while},%
  morekeywords=[3]{append,cap,close,complex,copy,delete,%
      len,make,new,panic,print,println,recover},%
  morekeywords=[2]{bool,map,byte,complex64,complex128,float32,float64,%
      int,int8,int16,int32,int64,rune,string,interface,struct,%
      uint,uint8,uint16,uint32,uint64,uintptr,chan,error,any},%
  keywordstyle=[1]{\bfseries\color{black}},
  keywordstyle=[2]{\bfseries\color{black}},
  keywordstyle=[3]{\bfseries\color{black}},
  commentstyle=\color{gray},
  backgroundcolor=\color{white},
  escapechar={@},
  moredelim=**[is][\color{darkgreen}+]{+++}{+++},
  moredelim=**[is][\color{red}-]{---}{---},
  numbersep=5pt,
  xleftmargin=10pt,
  numbers=left
}%

\lstdefinelanguage{CustomPyton}{%
aboveskip=5pt,
belowskip=0pt,
lineskip= {-1.5pt},
language=python,                %
basicstyle=\scriptsize,       %
numbers=left,                   %
numberstyle=\tiny,      %
stepnumber=1,                   %
numbersep=2pt,                  %
backgroundcolor=\color{white},  %
showspaces=false,               %
stringstyle=\scriptsize,
identifierstyle=\scriptsize,
commentstyle=\scriptsize,
basicstyle=\scriptsize\ttfamily,
showstringspaces=false,         %
showtabs=false,                 %
frame=tb,                   %
tabsize=2,                      %
captionpos=b,                   %
breaklines=true,                %
breakatwhitespace=false,        %
title=\lstname,                 %
  sensitive,%
  morecomment=[s]{/*}{*/},%
  morecomment=[l]//,%
  morestring=[b]',%
  morestring=[b]",%
  morestring=[s]{`}{`},%
  commentstyle=\color{gray},
  backgroundcolor=\color{white},
  escapechar={@},
  numbersep=5pt,
  xleftmargin=10pt,
  numbers=left,
  morekeywords=[1]{if,for,def,range,retrun,},%
  keywordstyle=[1]{\bfseries\color{black}},
}%

\begin{document}

\date{}

\title{Uber's Failover Architecture: Reconciling Reliability and Efficiency \\
in Hyperscale Microservice Infrastructure}

\author{
{\rm Mayank Bansal, Milind Chabbi, Kenneth Bøgh, Srikanth Prodduturi, Kevin Xu, Amit Kumar, David Bell,}\\
{\rm Ranjib Dey, Yufei Ren, Sachin Sharma, Juan Marcano, Shriniket Kale, Subhav Pradhan,}\\
{\rm Ivan Beschastnikh$^\dagger$, Miguel Covarrubias, Chien-Chih Liao, Sandeep Koushik Sheshadri, Wen Luo,}\\
{\rm Kai Song, Ashish Samant, Sahil Rihan, Nimish Sheth, Albert Greenberg, Uday Kiran Medisetty}\\
Uber Technologies \ \ \ $^\dagger$University of British Columbia\\
\textit{\{mabansal, milind, bgh, sproddut, kevinxu, amitkr, dastbe, ranjib, yufeir, sachins\}@uber.com}\\
\textit{\{marcano, skale, subhav, miguel.covarrubias, ccliao, sandeep.sheshadri\}@uber.com}\\
\textit{\{wenl, kaisong, asamant, rihan, nimish, albert.greenberg, udaym\}@uber.com, \ $^\dagger$bestchai@cs.ubc.ca}
}

\maketitle

\begin{abstract}

Operating a global, real-time platform at Uber's scale requires infrastructure that is both resilient and cost-efficient.
Historically, reliability was ensured through a costly 2$\times$ capacity model---each service provisioned to handle global traffic independently across two regions---leaving half the fleet idle. We present Uber's Failover Architecture (\ufa{}),
which replaces the uniform 2$\times$ model with a differentiated architecture aligned to business criticality.
Critical services retain failover
guarantees, while non-critical services opportunistically use failover buffer capacity reserved for critical services during steady state.
During rare ``full-peak'' failovers, non-critical services are selectively preempted and rapidly restored, with differentiated Service-Level Agreements (SLAs) using on-demand capacity.
Automated safeguards, including dependency analysis and regression gates, ensure critical services continue to function even while non-critical services are unavailable.
The quantitative impact is significant: \ufa{} reduces steady-state provisioning from 2$\times$ to 1.3$\times$, raising utilization from $\sim$20\% to $\sim$30\% while sustaining 99.97\% availability.
To date, \ufa{} has hardened over 4,000 unsafe dependencies, eliminated over one million CPU cores from a baseline of about four million cores.

\end{abstract}

\section{Introduction}
\label{sec:intro}

At hyperscale, distributed systems face a persistent tension between reliability and cost of capacity~\cite{borg2015, tailatscale2013}.
For Uber, where millions of earners rely on the platform for their livelihood and hundreds of millions depend on it for mobility and delivery services, reliability has always been paramount---even at the expense of significant redundancy in compute resources to ensure failure safety.
For years, Uber operated with a dual-region active-active model~\cite{Dynamo2007,spanner2012}, where each region could absorb 100\% of global traffic during a regional failure. While robust, this design effectively doubled our steady-state capacity requirements. 

Compounding the issue, our reliability strategy was homogeneous: every service, from critical trip-matching  to developer testing workloads, was provisioned with the same SLAs.
This uniformity simplified software design, development and operations but imposed significant and unnecessary costs.
The impetus for change came from a data-driven analysis of four years of failover events, which revealed a striking insight: catastrophic, full-peak failovers, the very events justifying the 2$\times$ model, were exceedingly rare, occurring less than 20 hours per year on average.
This finding provided a principled justification for abandoning our rigid, ``one-size-fits-all'' model and designing a more cost-effective architecture for handling failures.

Capitalizing on this insight, however, has been a formidable undertaking.
The primary challenge lay in the structural complexity of our microservices. A mesh of over 6,000 microservices deployed in over 16,000 environments had evolved with a tangled dependency graph~\cite{errorscopeLee, crispZhang}.
Critical, user-facing services had developed unsafe, ``fail-close'' dependencies on non-critical services, meaning the failure or absence of a non-critical service could cascade upwards and trigger a core business outage.
Some of this behavior was accidental, but some was because of a general practice of propagating downstream RPC failures up to their callers.
Furthermore, our platform lacked native mechanisms to safely differentiate workloads. Ensuring high availability of any service required the same level of availability for the transitive closure of its dependencies, irrespective of their business-criticality. As a result, all services were over-provisioned for failover.
These technical hurdles were magnified by a decentralized engineering culture, where hundreds of teams deploy services independently, making any top-down architectural change a massive socio-technical challenge.

To overcome these obstacles, we developed Uber's Failover Architecture (\ufa{}).
A key feature of \ufa{} is a differentiated, business-aligned availability model that replaces uniform SLA with tiered SLAs.
Under \ufa{}, only the most critical services retain instantaneous failover guarantees by maintaining a standby failover buffer.
These buffers, however, are not sitting idle.
A second key feature is an architecture for intelligent capacity oversubscription, where non-critical services are opportunistically scheduled onto the unused failover buffers of critical services.
During a rare peak regional failover, these less critical services are quickly but preferentially preempted from the failover buffers to make room for critical services.
The evicted, non-critical services are then rapidly restored within a strict but differentiated Recovery Time Objective (RTO) by provisioning them into elastic cloud capacity and/or by evicting batch jobs that can tolerate higher downtime.

This paper details the journey of \ufa{}, from its data-driven inception to its architectural design, phased implementation, and measured impact.
In summary, the key contributions of this work are:

\begin{itemize}[leftmargin=*,noitemsep, nolistsep]
\item Real data on hyperscale microservice compute infrastructure at Uber. 

\item An architecture for intelligent failure management via oversubscription and workload shedding: A novel architecture where non-critical workloads, during non-failure situations, run within idle failover buffer capacity designated for critical services, coupled with an automated system to shed and restore these workloads. 

\item A multi-layered safety and regression prevention strategy: A suite of analysis tools and an automated regression detection system in the CI/CD pipeline to proactively detect and prevent unsafe ``fail-close'' dependencies.

\item Large-scale deployment and evaluation on a production system of over 6,000 microservices: \ufa{} eliminated about one million CPU cores out of a baseline of about four million CPU cores, demonstrating a significant increase in resource utilization from $\sim$20\% to $\sim$30\%.
We also present a series of operational lessons learned.

\end{itemize}

\section{Background and Terminology}

\begin{figure}[!t]
    \centering
    \includegraphics[width=.9\linewidth]{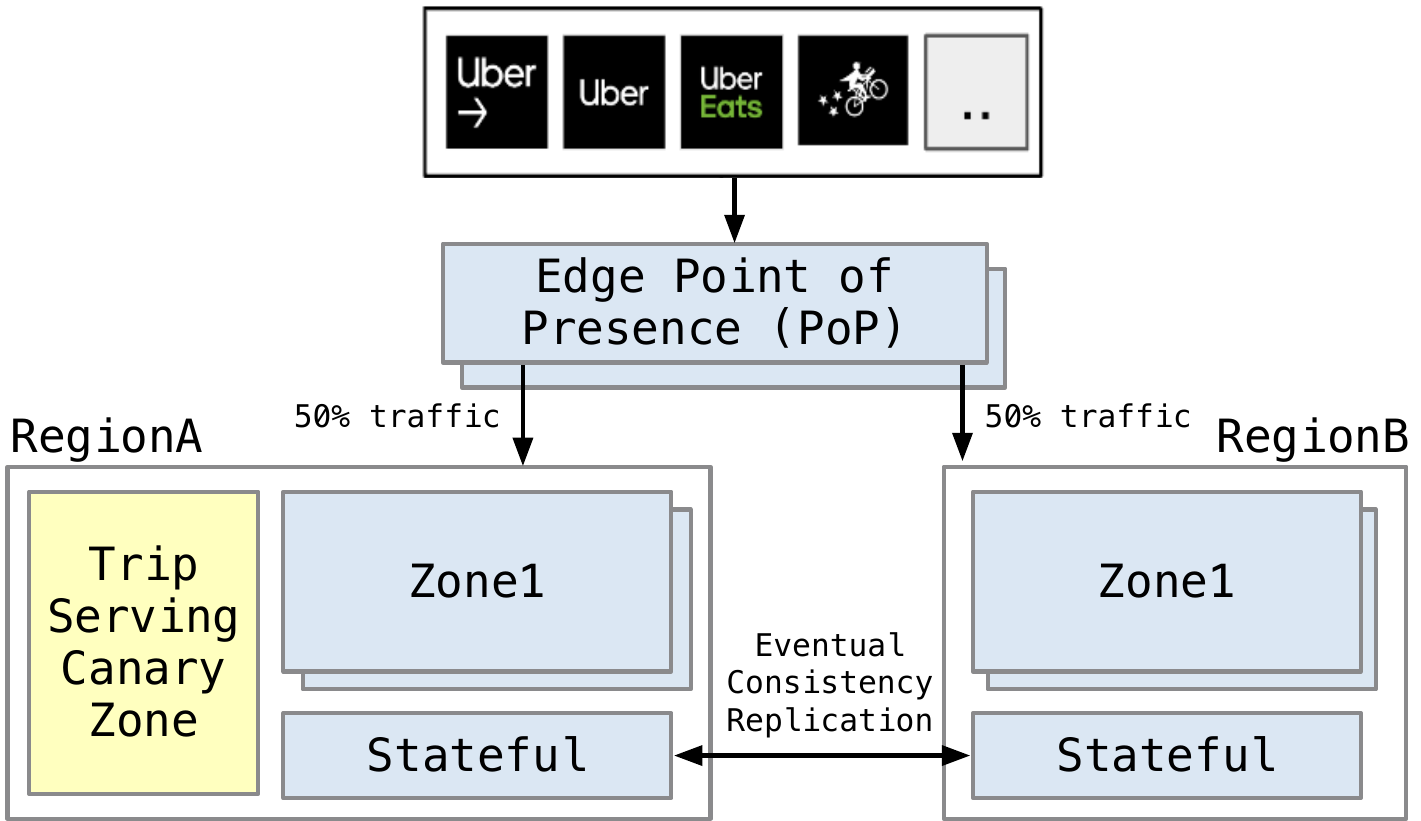}
    \caption{Legacy architecture (pre-\ufa{}).}
    \label{fig:legacy}
\end{figure}

Uber's platform is built on a dual-region active-active microservice architecture depicted in Figure~\ref{fig:legacy}.
Understanding the scale, complexity, and inherent inefficiencies of this architecture is essential context for UFA.

Uber's microservices are deployed in two regions, or geographically distributed data centers.
Each region comprises multiple zones spread both on-premise (capital expenses model) and in the cloud (operating expenses model).
Each region comprises stateless and stateful services.

During normal operations (steady state), each city's traffic is routed to exactly one region.
However, every microservice is provisioned with enough compute resources (CPU, memory, and disk) to be able to handle a regional failover, i.e., when one region becomes unavailable due to issues such as software faults, hardware problems, or large-scale infrastructure disruptions, while the other region continues to operate, absorbing extra traffic.
Since a failure is to be expected at any time, each region is provisioned with twice the peak expected capacity.
A failure is treated as \emph{peak} if the number of users on a trip is more than 85\% of the weekly peak; non-peak otherwise.
A failure is considered \emph{full} if more than 50\% of the cities fail over; partial otherwise.
A \emph{full-peak} failover happens when both conditions are true.

\textbf{Scale and service tiering:}  Uber operates approximately 6,000 microservices across 16,000 environments (e.g., prod, staging, canary), with services organized into business‑criticality tiers T0 (higher-priority aka critical) to T5 (lower-priority aka non-critical) as shown in Table~\ref{tab:tiersandcores}.
Non-production represents a special tier where services are not directly involved in production but may receive test or shadow traffic from production.
Prior to UFA, the baseline steady‑state allocation was about 2 million CPU cores per region and 4 million globally.
These figures provide context for the magnitude of the fleet and the cost of treating all tiers as equally mission-critical.

\begin{table}[!t]
\scriptsize
    \centering
    \begin{tabular}{c|l|r}
        \textbf{Tier} & \textbf{Description} & \textbf{Baseline \#  CPUs} \\\hline
        T0 & Infrastructure and critical applications & 201K \\
        T1 & Critical trip flow & 3.03M \\
        T2 & Business critical applications & 400K \\
        T3 & Internal tools critical to applications & 254K\\
        T4 & Internal tools used by Employees  & 23.1K \\
        T5 & Test versions and the rest  & 22.1K \\
        NP & Non-production (staging, shadow, etc) & 249K \\
    \end{tabular}
    \caption{Tiers, their semantics, and baseline core counts.}
    \label{tab:tiersandcores}

\end{table}

The complexity of service interactions is immense,
as shown in Table~\ref{tab:traffic}, which captures cross-tier remote-procedure calls (RPCs) over a randomly selected week in 2025. The system handles about 62 trillion total requests in a week.
Notably, 32 trillion of these RPCs ($\sim$50\%) flow from a higher-priority tier to a lower-priority tier.
This is especially relevant to our work, as we later demonstrate the significant risks posed by these ``tier-inverted'' dependencies.
Each service exposes one or more endpoints callable via RPC, and Table~\ref{tab:endpoints} provides statistics on the number of unique endpoints per tier, along with their percentile distribution, which illustrates the system's scale and diversity.

\begin{table*}[!t]
\begin{minipage}{0.63\textwidth}
\scriptsize
    \centering
    \begin{tabular}{|c|c|c|c|c|c|c|c|c|}
    \hline
    & \multicolumn{7}{c|}{\textbf{Callee Tier}} & \\
\textbf{caller}	&\textbf{T0}	&\textbf{T1}	&\textbf{T2}	&\textbf{T3}	&\textbf{T4}	&\textbf{T5} &\textbf{NP} & \textbf{to lower-tier}\\ \hline
T0	&47.1B	&940B	&2.30T	&1.82T	&144B	&100B & 1.77T & 7.07T\\
T1	&10.7B	&21.8T	&2.24T	&387B	&6.07B	&70.4B & 18.6T & 21.3T\\
T2	&25.3B	&2.02T	&663B	&77.0B	&30.9M	&1.17B & 2.70T & 2.78T \\
T3	&7.95B	&288B	&119B	&16.9B	&192M	&6.09B & 1.06T & 1.07T \\
T4	&788M	&11.5B	&599M	&228M	&1.19B	&12.1M & 22.1B & 22.1B\\
T5	&290M	&76.1B	&266M	&849M	&1.30M	&4.52B &  14.1B & 14.1B \\
NP	&107B	&1.53T	&471B	&126B	&12.8B	&18.3B	& 3.13T & 0\\
\hline
    \end{tabular}
    \caption{Cross-tier requests during a week in 2025. M=Million, B=Billion, T=Trillion.}
    \label{tab:traffic}
\end{minipage}\hfill
\begin{minipage}{0.34\textwidth}
\scriptsize
    \centering
    \begin{tabular}{|r|r|r||r|r|r|}
    \hline
    & & & \multicolumn{3}{c|}{\textbf{Endpoints/Service}} \\
\textbf{tier}&	\textbf{\#services}	& \textbf{\#endpoints}	& \textbf{p50}	& \textbf{p90} &   \textbf{max} \\\hline
T0&	96&	1452&		7&		79&		153 \\
T1&	607&	17643&		10&		96&		1416 \\
T2&	561&	12362&		11&		82&		629 \\
T3&	1550&	15255&		3&		38&		1059 \\
T4&	283&	1672&		2&		22&		116 \\
T5&	882&	5170&		2&		13&		1953 \\
NP & 	18K	& 77K&		1&		18&		1594 \\ \hline

    \end{tabular}
    \caption{Statistics of unique endpoints/tier observed during nine months in 2024/25.}
    \label{tab:endpoints}
\end{minipage}
\end{table*}

Next, we describe the infrastructure components needed to understand the rest of the paper.

\begin{itemize}[leftmargin=*,noitemsep, nolistsep]

    \item \textbf{Up}~\cite{upwww}: Uber's service deployment platform. Up places and migrates services across federated Kubernetes~\cite{k8swww} (k8s) clusters in multiple zones in both regions. Deployment granularity is at the service-environment pair, where an environment may be production, canary, non-production, etc. Up orchestrates about 700K such deployments per week.

    \item \textbf{Compute}~\cite{computewww}: Uber's container orchestration layer, built atop a customized Kubernetes with Uber-specific control plane, resource definitions, and service discovery.
    Compute manages the low-level scheduling and resource allocation for services, launching over a million pods~\cite{k8spods} per day and handling instantaneous scale-ups during failover.

    \item \textbf{Batch clusters}:
Computational resources on clusters dedicated to non-real-time workloads (analytics and ML training) using engines powered by Spark~\cite{sparkwww}, Presto~\cite{prestowww} and managed by orchestrators like YARN~\cite{yarnwww} and Kubernetes.

    \item \textbf{Canary zone:}  A designated zone where every new deployment begins before proceeding to the remaining zones. It receives 2\% of regional traffic.

    \item \textbf{Outage Mitigator (OMG):} A tool to orchestrate region‑to‑region city migrations during failovers.
    In the legacy system, OMG was semi‑automatic and limited to city migrations. It was not capable of the orchestrations required for UFA failover/failback.
    As a result, failover/failback could be slow, risky, and operationally intensive.
\end{itemize}

In this paper, a region is in \emph{failover} state if it is serving traffic from a different primary region, which has been determined to be unhealthy.
We use the term \emph{failback} to mean the process of recovering from a failure by moving service traffic back from the failover region to the original, now-healthy, region.

\smallskip
\noindent\textbf{Scope.} \ufa{} targets Uber's \emph{stateless} microservices. These constitute the vast majority of compute capacity shown in Table~\ref{tab:tiersandcores} and account for about 75\% of compute cores.
Extending \ufa{}'s differentiated model to stateful services (e.g., databases, caches) is future work.

\label{sec:background}

\section{Problems, Key Insights, and Constraints}
\label{sec:motivation}

\begin{figure}[!t]
\centering
\begin{minipage}[b]{.6\linewidth}
    \centering
    \includegraphics[width=\linewidth]{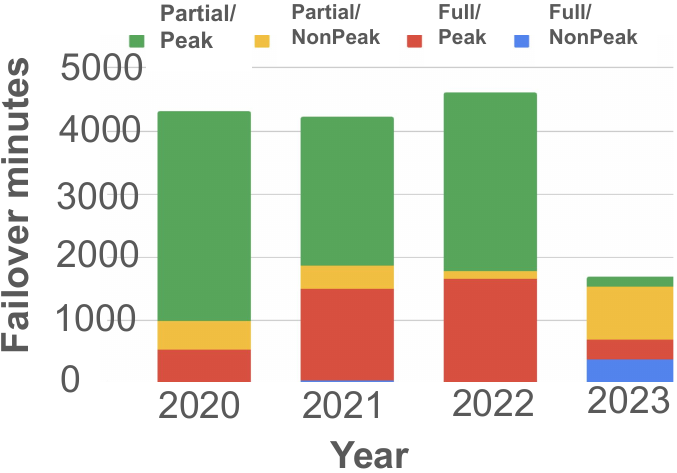}
    \caption{Full-peak failover minutes is a small (0.23\%) fraction of time in a year and on a reduction trend.}
    \label{fig:failureminutes}
\end{minipage}%
 \hfill
 \begin{minipage}[b]{0.35\linewidth}
     \centering
\scriptsize
        \begin{tabular}{c|c}
            \textbf{Year} & \textbf{\# regional failovers} \\             \hline
            2020 & 25\\
            2021 & 24\\
            2022 & 16\\
            2023 & 13\\
            \hline
        \end{tabular}
        \caption{Yearly count of regional failovers.}
        \label{tab:failovercnt}

 \end{minipage}
\end{figure}

The existing architecture faced two critical problems:

\smallskip

\noindent \textbf{Problem 1: Low CPU utilization.} Identical SLA for all tiers leaves half the fleet idle at steady state.
Analysis of multi-year incident data before \ufa{} (Figure~\ref{fig:failureminutes}) showed that full-peak regional failovers average fewer than 20 hours annually, just 0.23\% of the time\footnote{The spike in failover minutes from 2020 to 2021 was an anomaly caused by the COVID-19 slump in 2020.}.
The frequency of failovers (Figure~\ref{tab:failovercnt}) also steadily declined between 2020 and 2023.
Such rarity means most failover buffers sit idle almost always.
The current model depresses the average fleet-wide CPU utilization down to $\sim 20\%$.

\smallskip

\noindent \textbf{Problem 2: Fail-close outages:} Over a decade of operations shows that even a highly over-provisioned architecture is not immune to critical outages.
A common root cause is the dependency between critical and non-critical services: when the latter fail, critical services could degrade too, violating their SLAs.
Critical services are typically built and operated with stricter safety and isolation standards, while less-critical services are not.
This asymmetry becomes risky when dependencies---direct or transitive, intentional or accidental---emerge across tiers.
Failures in a less critical service, whether due to hardware faults, bugs, misconfigurations, or overload, cascade into critical service outages.
Identifying and eliminating such cross-tier fail-close dependencies is both a technical and organizational challenge, echoing dependency-safety concerns seen in other hyperscale systems~\cite{capa2024}.

\paragraph{Low CPU utilization can be improved through capacity sharing, but only \emph{after} eliminating fail-close dependencies.}
Low CPU utilization enables capacity \emph{oversubscription}: lower-priority services can run opportunistically in the reserved headroom of higher-priority services during steady state~\cite{Heracles2015}.
During a failover, we can swiftly evict lower-priority services to make room for the continued operation of higher-priority services.
Crucially, such eviction is only safe once cross-tier \emph{fail-close} edges are eliminated, so that critical services \emph{fail open} (do not propagate downstream failures to upstream) when lower-priority services are unavailable.

\noindent
\textbf{One program, two goals.}
Systematically fixing cross-tier fail-close dependencies and introducing differentiated eviction and reinstatement of lower-tier services during a failover \emph{solves both problems at once}. It turns idle buffers into useful capacity in steady state (improving utilization) \emph{and} reduces outage blast radius by removing unsafe dependencies.
This insight is at the core of the \ufa{} program.

\noindent
\textbf{Constraints to ensure safety.}
Safe adoption of the \ufa{} model requires:
\begin{itemize}[leftmargin=*,noitemsep, nolistsep]
    \item \textbf{Tiered reliability.} Critical services must retain sub-second Recovery Time Objective (RTO) in all scenarios; non-critical services should tolerate up to one hour RTO only in full-peak failovers.
    \item \textbf{Dependency isolation.} Critical services that survive during a failover must not have fail-close dependencies on services that may be preempted.
    \item \textbf{Resource bounds.} System oversubscription must remain within safe operating conditions validated by prior studies.
    \item \textbf{Elastic recovery.} Sufficient compute capacity must be available to restore evicted services within their SLAs.
    \item \textbf{Automated workflow.} Failover workflows must be fully automated to reclaim and repurpose buffers within minutes.
\end{itemize}

\textbf{The goal state:} Reduce steady-state capacity from $2\times$ to $1.3\times$, raising host utilization above 30\%, while preserving reliability for critical services and ensuring graceful degradation and recovery of non-critical services.

\textbf{Ruling out an obvious solution:}
An obvious solution would be to dedicate only steady-state compute capacity and elastically expand into the cloud during failovers in a ``pay-for-use'' model~\cite{disasterrecoverycloud2010}. This approach, however, breaks down at Uber's scale, where several hundred thousand to a million CPU cores are needed within a matter of seconds to minutes, which none of the large-scale cloud providers can guarantee.
Large cloud providers such as AWS~\cite{awswww}, Azure~\cite{azurewww}, GCP~\cite{gcpwww}, and OCI~\cite{ociwww} apply regional and/or per-account quotas on the number of CPUs. Scaling beyond these default limits requires quota increase requests, which is often a semi-manual process taking several days. 

In Section~\ref{sec:design}, we describe how \ufa{} operates during a failover, assuming all fail-close dependencies are eliminated.
In Section~\ref{sec:tools}, we describe the mechanism we put in place to detect and eliminate fail-close dependencies.

\section{The \ufa{} Design \& Architecture}
\label{sec:design}

\begin{table}[!t]
\scriptsize
    \centering
    \begin{tabular}{r|l|l|l}
       \textbf{Failure Class}  & \multicolumn{2}{|c|}{\textbf{On failover}}  & \textbf{RTO}  \\ \cline{2-3}
       & \emph{Container} & \emph{Network} &  \\ \hline
        \textbf{\alwayson{}} & In-place expand &  Uninterrupted & secs    \\
        \textbf{\activemigrate{}} & Live but migrated to batch & Reconfigured & secs   \\
        \textbf{\restorelater{}} & Stop and revive in batch/cloud & Brief Downtime & 1hr     \\
        \textbf{\terminate{}} & Stop and no restore & Downtime &  none
    \end{tabular}
    \caption{Failure profile classes and their failover behaviors.}
    \label{tab:failureprofile}
\end{table}

\begin{figure}[!t]
  \centering
\includegraphics[width=.8\columnwidth]{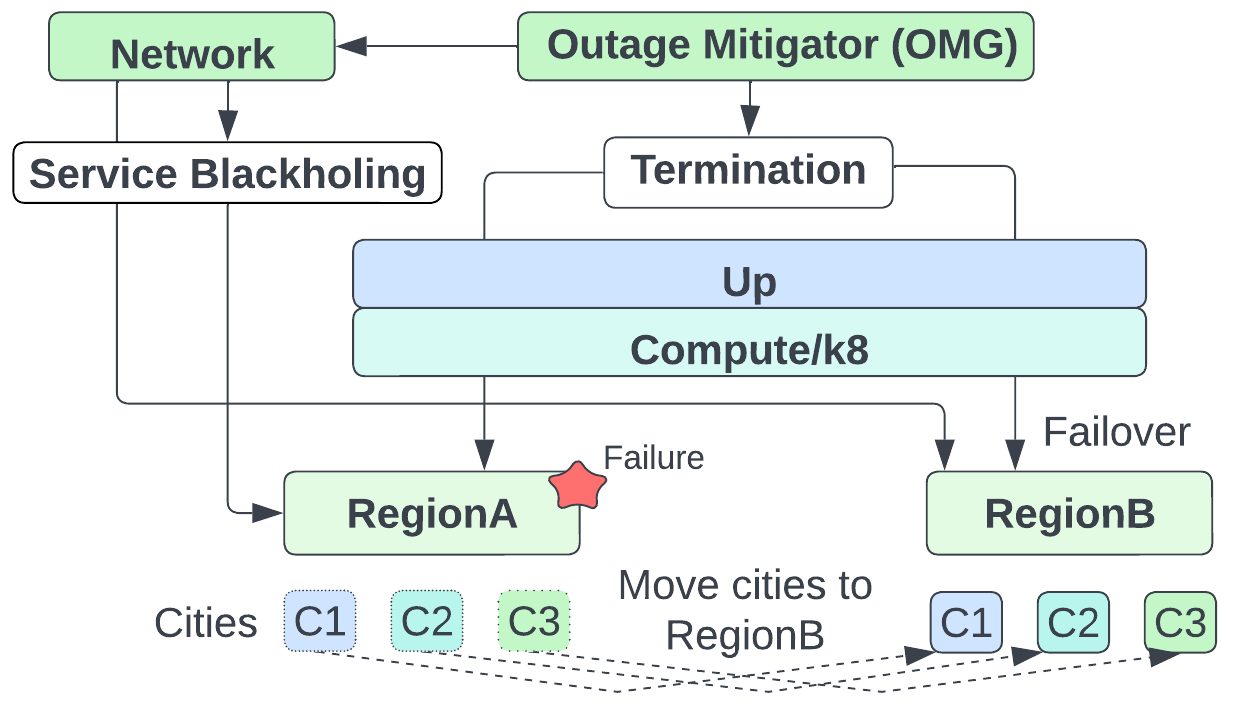}
  \caption{Key components involved in handling a failover from RegionA to RegionB.}
  \label{fig:arch}
\end{figure}

\begin{figure}[t]
  \centering
\includegraphics[width=\columnwidth]{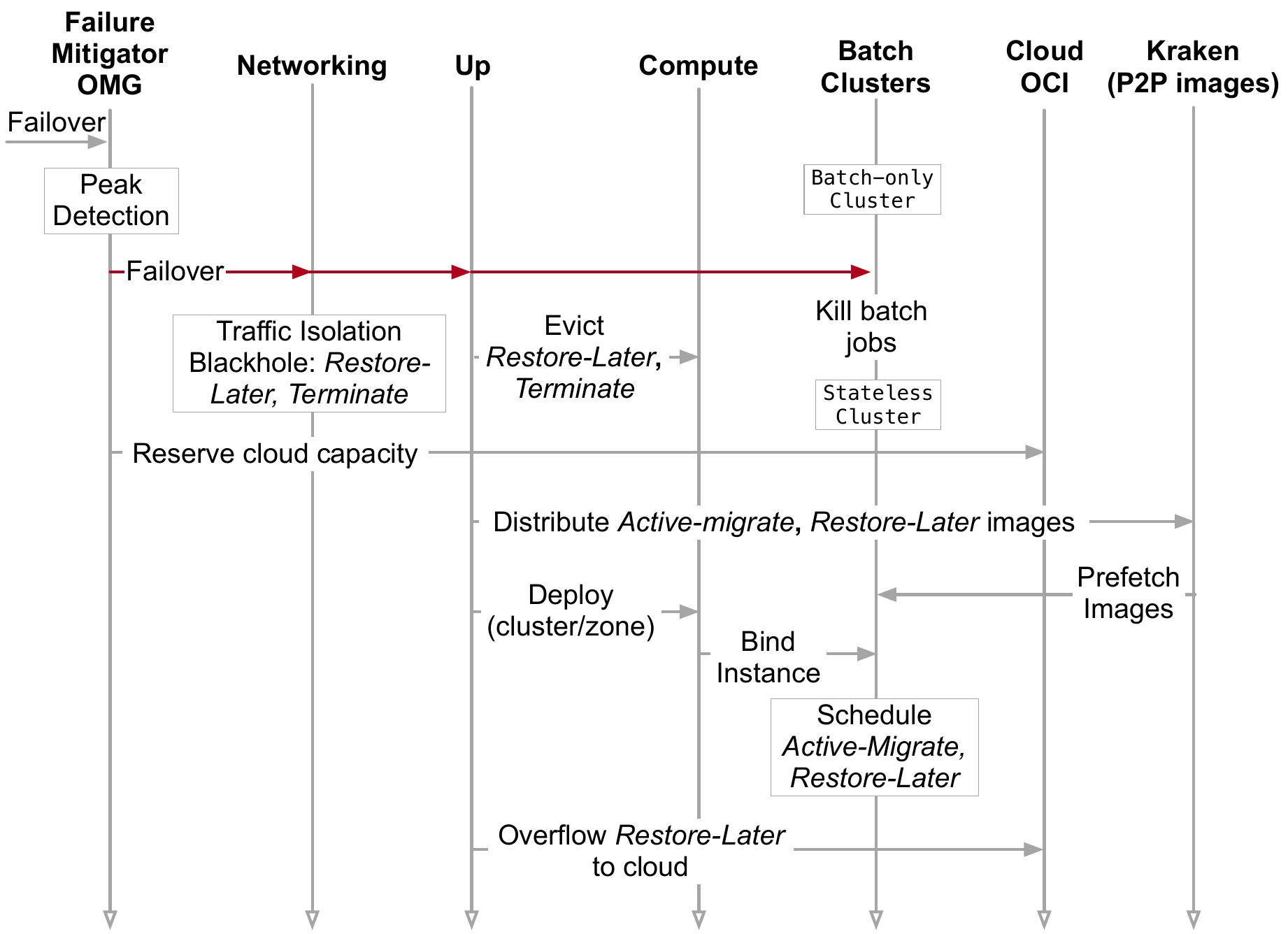}
  \caption{Events coordinated during a \ufa{} failover.}
  \label{fig:failover}
\end{figure}

\ufa{} classifies services into four failover behavior classes based on their behavior during a failover, shown in Table~\ref{tab:failureprofile}.
The \textbf{\alwayson} class services are not preempted during a failover; they scale up into their dedicated failover CPU buffers; typically, T0 and T1 services are in this class.
The \textbf{\activemigrate{}} class services are brought up elsewhere before being terminated from their original hosts, ensuring continuous availability; typically, T2 services are in this class.
The \textbf{\restorelater} class services are instantly terminated and restored later elsewhere, with an up to 1-hour service disruption; all T3-T5 services are in this class.
The \textbf{\terminate} class services are instantly terminated, causing interruption until they are restored after failback; all non-production (NP) services are in this class.
The mechanism determines where to restore terminated services: \ufa{} repurposes batch clusters first, then the cloud.
We refer to the compute clusters that host services during normal operation as ``steady-state'' clusters, and to batch clusters that are repurposed during a failover to host evicted services as ``burst'' clusters.

In a nutshell, the goal during a peak failover is to free enough capacity in the surviving region for \alwayson{} services to absorb $2\times$ traffic, while preserving \activemigrate{} availability and restoring \restorelater{} services within one hour.
The process, orchestrated by the Outage Mitigator (OMG), proceeds in order shown in Figure~\ref{fig:failover}:
\emph{(1)} \terminate{} and \restorelater{} services are immediately evicted from steady-state clusters, freeing CPU headroom for \alwayson{} services to scale up.
\emph{(2)} Batch clusters are simultaneously converted into ``burst'' clusters by evicting batch jobs and prefetching container images for fast launch.
\emph{(3)} \activemigrate{} service instances are brought up in the burst clusters, network traffic is redirected to them, and their old instances in the steady-state cluster are terminated. This progressively makes room for \alwayson{} to expand. This repeats city by city until the entire $2\times$ traffic is served.
\emph{(4)} Concurrently, \restorelater{} services are restored in batch and/or cloud capacity within a 1-hour RTO.

Next, we detail each component involved in this workflow (Figure~\ref{fig:arch}):
failover orchestration (\S~\ref{sec:omg}),
networking and traffic isolation (\S~\ref{sec:networking}),
service re-deployment (\S~\ref{sec:up}),
resource partitioning and oversubscription (\S~\ref{sec:compute}),
batch conversion and cloud bursting (\S~\ref{sec:batch}, \S~\ref{sec:cloud}),
and the recovery process (\S~\ref{sec:recovery}).

\subsection{Failover Orchestration with OMG}
\label{sec:omg}

OMG is the \ufa{} failover orchestration entry point.
To support \ufa{}, existing failover and failback workflows in OMG were changed to \emph{automatically} (a) detect mode of operation, (b) ensure availability of burst capacity, and (c) migrate services between steady-state and burst clusters.

A key feature of the orchestrator is detecting the failover mode, which can be either \textbf{peak} or \textbf{non-peak}. This is done based on: (a) the peak traffic volume observed during the past week ($tv_{peak}$), (b) the traffic volume at failover time ($tv_{failover}$), and (c) a threshold that is periodically computed based on the failover multiplier ($T$). The mode of a failover will be peak if $tv_{failover} \geq T \times tv_{peak}$. 

This mode detection is critical because, in non-peak mode, both the failover and failback workflows are responsible only for moving city traffic between regions. However, in peak mode, both of these workflows are more complex.

During a peak failover, the orchestrator begins by locking down services of all failure classes other than \alwayson{}, to prevent changes during the process. It then ensures sufficient burst capacity by preparing on-prem batch clusters to host bursted services, while dynamically provisioning cloud capacity if batch resources prove insufficient. Next, it disables services in the \restorelater{} and \terminate{} failure classes within the steady-state cluster, which involves throttling their traffic and suppressing alerts. The orchestrator then migrates services in the \restorelater{} and \activemigrate{} classes to the burst cluster. Note that the \terminate{} class services do not qualify and therefore remain disabled throughout a failover. Finally, it completes the failover by enabling \restorelater{} services that successfully respawn in the burst cluster and moving traffic from the source region to the destination region.

During peak failback, the orchestrator first moves traffic back to the source region. It then initiates the migration of all previously bursted services back to the steady-state cluster. It also enables \terminate{} class services in the steady-state cluster, and finally, it unlocks all previously locked services.

\subsection{Networking During Failover}
\label{sec:networking}

\ufa{}'s differentiated SLA model requires precise traffic isolation: \alwayson{} and \activemigrate{} services must maintain connectivity while \restorelater{} and \terminate{} services are blocked during failovers.
This isolation serves two purposes: (1) allowing critical services to gracefully handle the absence of non-critical dependencies, and (2) preventing thundering herd~\cite{Mogul1997} effects when \restorelater{} services are terminated and later restored.
Traffic isolation is enforced through two complementary mechanisms.

\emph{First}, Uber's L7 service mesh coordinates hundreds of thousands of client-side load balancers using look-aside load balancing and rate limiting.
For \ufa{}, we augmented the rate-limiting system to apply cross-cutting policies based on caller and callee.
To limit the blast radius of these new policies, we added zonal and regional isolation to the configuration pipeline, while retaining global convergence within approximately 30 seconds.

\emph{Second}, Uber operates a security overlay with per-workload access controls.
We extended it to support ``global policies'' that block all connectivity into \restorelater{} and \terminate{} services during failover, and re-enable it once those services are restored.
This ensures fail-fast behavior even for peer-to-peer traffic that bypasses the load balancers.

The service mesh also provides observability during failovers: blocked requests carry error messages identifying the cause, and failover-related events are tagged so that alert and SLA calculations exclude them, avoiding noise for service owners.

\subsection{Rapid Service Re-deployment with Up}
\label{sec:up}

During a failover, Up uses two migration strategies: break-before-make (BBM) for \restorelater{} services, which tolerate brief downtime, and make-before-break (MBB) for \activemigrate{} services requiring zero downtime.

During \ufa{} failovers, Up coordinates a multi-phase orchestration process.
Upon receiving a ``preheat'' signal from OMG, Up triggers Docker image prefetching into burst datacenter zones and signals batch clusters to evict preemptible workloads, preparing capacity for incoming stateless services.
When the failover signal arrives, Up executes immediate termination of \restorelater{} and \terminate{} class service environments across all steady-state clusters, bypassing the normal workflow overhead by directly signaling cluster-level kills.
Simultaneously, Up begins tier-by-tier migration of \activemigrate{} services, moving up to 2,000 environments in parallel with dynamic routing reconfiguration. Up's massive parallelism is enabled by Uber's asynchronous workflow engine~\cite{cadenceUber}.

A key enabler is image prefetching via a peer-to-peer Docker registry~\cite{krakenUber}.
On-demand image pulls during mass migrations cause severe bottlenecks due to origin server saturation and kubelet~\cite{kubeletwww} backoff.
By proactively downloading several terabytes of image data into burst zones during batch conversion, Up reduces service startup time by up to 30\%. 

Beyond failover, Up continuously manages service eligibility for \ufa{} through an automated reconciliation system that onboards and offboards services.
It validates service tiers, detects fail-close dependencies, checks hardware requirements, and enforces deny-list policies. Services with special needs (e.g., GPUs) are automatically excluded, while new services are onboarded after a seven-day stabilization period.

\subsection{Smart Over-Subscription with Compute}
\label{sec:compute}

The Compute subsystem is the core of \ufa{}'s intelligent oversubscription~\cite{BorgOmegaK8CACM2016, borg2015}. 
While Up orchestrates the high-level service migrations, Compute handles the low-level resource partitioning, workload co-location, and rapid failover execution.

\textbf{CPU Resource Partitioning and Oversubscription.} Each host's CPU capacity is partitioned into two resource pools: \texttt{stateless.cpu} for \alwayson{} services and \texttt{overcommit.cpu} for differentially preemptible workloads.
This is implemented via Kubernetes extended resources: a node agent advertises additional CPU capacity based on cluster configuration. For example, a host with 100 physical CPUs advertises 150 total: 100 in the stateless pool and 50 in the overcommit pool.
The Kubernetes scheduler treats these as separate resource types, so overcommit pods cannot interfere with critical service placement.

A key challenge is determining the safe overcommit factor for a Kubernetes cluster~\cite{AutothrottleNSDI2024}.
From Table~\ref{tab:tiersandcores}, the overcommit pool must cover $\sim$1M cores,
smaller than the 3M cores reserved for \alwayson{} services, making it the limiting factor and a source of fragmentation when allocating services to host pools.
We built a simulator that models cluster configurations and workloads, which recommended a $1.5\times$ overcommit factor.
This 50\% headroom is allocated to preemptible services.
The core-to-memory ratio further constrains oversubscription.
Let $M_h$ = average memory per host core, $M_s$ = memory per service core,
$\alpha_m$ = safe memory allocation fraction, and $\alpha_c$ = safe CPU allocation fraction.
Then, $O_{\max} = \frac{M_h}{M_s} \times \frac{\alpha_m}{\alpha_c}$, is the maximum achievable overcommit.
In our clusters, $M_h = 8$ GB/core, $M_s = 4$ GB/core, $\alpha_m = 0.75$, and $\alpha_c = 0.9$, yielding a theoretical limit of $O_{\max} = 1.66\times$.
$\alpha_m$ and $\alpha_c$ are deliberately set below $1.0$ to account for system overheads. Specifically, this ensures that a portion of resources is reserved for host agents.
It also maintains spare capacity, which is essential for operations.

\textbf{Workload Prioritization and Performance Isolation.} To ensure that overcommitted workloads do not degrade critical service performance, Compute implements differentiated resource allocation through Linux cgroups.
Services are assigned CPU shares and limits based on their tiers, with \alwayson{} services receiving high-priority scheduling
while \restorelater{} and \terminate{} services receive dramatically reduced shares.
When contention occurs, less-critical services are automatically throttled while critical services maintain their performance guarantees.

\textbf{Rapid Failover Execution and Safety Mechanisms.} During failovers, Compute coordinates with Up to execute the BBM and MBB strategies described above. A failover controller terminates overcommit workloads immediately by setting replica counts to zero, bypassing normal Kubernetes workflows for speed.
Three safety mechanisms complement this: (1) a QoS controller evicts pods when node CPU utilization exceeds 75\%, cooling nodes to below 70\% utilization; (2) a utilization-aware scheduler considers current host utilization during placement; and (3) name discovery safety circuits are temporarily disabled to allow mass service discovery cleanup.

Co-hosting \alwayson{} and \activemigrate{} services could cause CPU surges during failover, but three factors prevent this.
First, traffic is migrated in city-sized groups with stability checks between migrations, providing time to spin up MBB containers before load increases.
Second, hosts are provisioned with a mix of all four classes, so evicting \terminate{} and \restorelater{} services frees enough CPU to absorb short-term spikes.
Third, if host-level CPU utilization still exceeds safe thresholds, QoS agents throttle or relocate \activemigrate{} instances.

\subsection{Batch Conversion}
\label{sec:batch}
During a failover, \ufa{} acquires new capacity by converting existing batch clusters into burst clusters to host \activemigrate{} and \restorelater{} services. When the Compute Capacity Orchestrator receives a failover signal from Up, it triggers the termination of preemptible (lower priority) batch workloads.
The workloads are evicted
and the hosts are prepared for stateless services.
This process is significantly faster than provisioning new machines.
Once the batch cores are converted to a burst cluster, the orchestrator announces the new stateless capacity to Up.
Up then deploys the evicted services into the new capacity.
An asynchronous spawner component provisions up to 240K CPUs across 2,000 hosts in under 20 minutes, enabling \ufa{} to re-establish thousands of service environments in parallel. 

\subsection{Cloud Bursting as Last Resort}
\label{sec:cloud}

Cloud bursting serves as a final safety net for restoring evicted \restorelater{} services within their one-hour RTO. 
Despite its external dependencies and cost, this capability has transformed a historically manual, multi-day process into an automated, sub-60-minute operation.
The workflow is tightly integrated with the failover orchestration and is only triggered when internal burst capacity is estimated to be insufficient to restore all \restorelater{} services.
When activated, the orchestrator rapidly provisions tens of thousands of cores across multiple cloud zones while adhering to quotas and availability constraints.
We developed this asynchronous provisioning workflow in collaboration with cloud providers to ensure rapid and reliable resource acquisition.

\subsection{Recovery Process}
\label{sec:recovery}

\begin{figure}[!t]
  \centering
\includegraphics[width=\columnwidth]{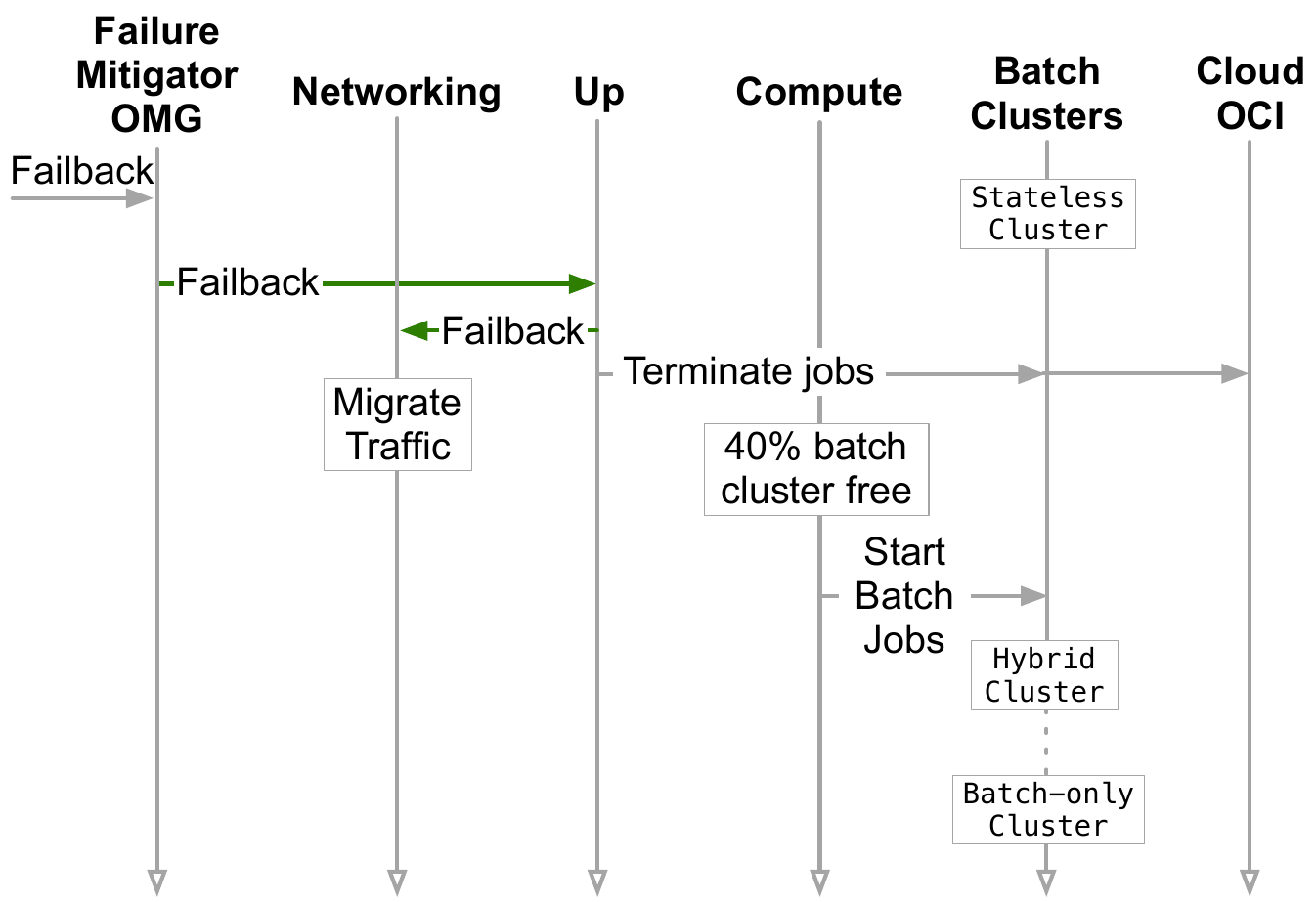}
  \caption{Failback process.}
  \label{fig:failback}
\end{figure}

The \ufa{} failback (recovery) process (Figure~\ref{fig:failback}) mirrors the failover MBB workflow. In particular, the same components are involved, and the emphasis is on releasing cloud and/or batch resources.

An operator manually triggers failback through the orchestrator once the failed region is confirmed as healthy.
During failback, traffic is moved to the now-operational region, one city at a time.
First, we shift services into the steady-state clusters.
We then update the traffic rules to route traffic to the new instances.
Finally, we terminate the previous instances.
We wait for 40\% of the batch capacity to be freed up before we allow the batch cluster
to take on new batch jobs.
If the cloud is in use, cloud resources are released.

The failback process entails many other decisions~\cite{recoverygraphs2006} that we omit due to space constraints.

\section{Validation Through Systematic Drills}
\label{sec:drills}

Deploying \ufa{} at Uber's scale requires rigorous validation.
\ufa{} employs a comprehensive drill program to validate both individual component resilience and end-to-end failover orchestration through controlled production exercises.
Drills operate through two complementary validation strategies: dependency safety certification and \ufa{} failover certification.

Dependency safety certification validates the fail-open behavior critical to \ufa{}'s oversubscription model by incrementally blocking (blackholing) traffic to services using the traffic isolation described earlier.
Services progress through a graduated traffic blackholing from 0\% to 100\%, replicating the worst-case scenario in which \restorelater{} and \terminate{} class services become completely unavailable during a failover.
Only services that successfully maintain functionality under complete dependency isolation achieve dependency safety certification, ensuring they can operate safely when co-located with preemptible workloads.

\begin{table}[!t]
\scriptsize
    \centering
    \begin{tabular}{r|r}
 & Cores returned \\ \hline
\terminate{} class   & 263K \\
Tier4/5 \restorelater{} class & 62K \\
Tier3 \restorelater{} class  & 159K \\
Tier2+ \activemigrate{} class & 92K \\
Tier1+ \activemigrate{} class & 455K \\
    \end{tabular}
    \caption{Phased cores returned via readiness reviews.}
    \label{tab:coresreturnedviaPRR}
\end{table}

\ufa{} failover certification validates the end-to-end orchestration workflow, exercising the full sequence of traffic throttling, service termination, core scaling, cloud bursting, and failback operations described previously.
These drills are conducted at both peak traffic (when riders-on-trip exceeds 85\% of weekly peak) and non-peak traffic conditions, with all cities failed over to the destination region to replicate maximum stress scenarios.

To expand \ufa{}'s scope, Uber implemented a structured readiness review process to manage risk.
Each phase required stakeholder sign-off and detailed documentation of drill plans and risk mitigation strategies before proceeding.
The program progressed from non-production workloads being treated as a \terminate{} class to other failure profile classes, as shown in Table~\ref{tab:coresreturnedviaPRR}.

Each drill is monitored through endpoint latency, error rates, and throughput-per-core metrics, with post-drill analytics automatically comparing against pre-drill baselines to identify regressions.
This approach has enabled about 43 production drills and 70+ staging drills while maintaining production stability.
This structured review process resulted in returning over one million CPU cores in 11 months, as shown in Table~\ref{tab:coresreturnedviaPRR}.

\section{Tools for Ensuring Regression Safety}
\label{sec:tools}

\ufa{}'s capacity oversubscription depends on eliminating fail-close dependencies between critical (\alwayson{} and \activemigrate) and non-critical (\restorelater{} and \terminate{}) services.
We use a multi-layered approach~\cite{MicroserviceFaults2021, correlated2020, independenceAsService2014} combining runtime detection, static analysis, and end-to-end testing to identify and eliminate these unsafe dependencies before they reach production.

\textbf{Runtime Dependency Analysis.} The first layer monitors all live production traffic through Uber's RPC framework~\cite{yarpcUber}, analyzing a few trillion RPCs daily (Table~\ref{tab:traffic}) to detect fail-close dependencies.
The analysis correlates every request with its downstream failures: if a caller endpoint consistently returns errors when a callee endpoint fails, that dependency is classified as fail-close.
This provides comprehensive coverage that would otherwise require extensive manual testing or remain hidden until production failures.

\textbf{Static Analysis for Fail-Close Detection.} A complementary static analysis catches unsafe dependencies earlier by analyzing service source code before deployment.
This whole-service analysis traces call paths across Go and Java codebases to determine whether downstream RPC errors can propagate through the service to cause upstream failures. It analyzes error return patterns in Go and exception propagation in Java.
Each dependency is classified as fail-close or fail-open, revealing safety violations before code reaches production.

Table~\ref{tab:tierwisebugs} shows the breakdown of the total of 4,155 fail-close violations found by the runtime and static analysis tools, which helped ensure our services are resilient to failures. Section~\ref{sec:lessons} describes how teams resolved these violations in practice.
The runtime tool was developed first and had a longer lead time, exposing an initial large number of defects. In contrast, the static analysis tool, which was developed later, detected defects missed by runtime analysis in less commonly executed paths and also found issues during continuous integration (CI).
A full treatment of both analysis techniques is beyond the scope of this paper; our focus here is on their role in \ufa{}'s safety pipeline.

\begin{table}[!t]
\scriptsize
    \centering
    \begin{tabular}{r|r|r}
Tier inversion& Runtime& Static\\ \hline
\alwayson{}$\longrightarrow$\restorelater{} & 1,120 & 337 \\
\activemigrate{}$\longrightarrow$\restorelater{} & 1,757 & 777 \\
*$\longrightarrow$\terminate{} & 164 & - \\ \hline
Total=4,155 & 3,041 & 1,114\\
    \end{tabular}
    \caption{Number of fail-close violations found by runtime and static analysis across different reliability classes.}
    \label{tab:tierwisebugs}
\end{table}

\textbf{The Canary Regression Gate.}
The canary regression gate is an automated regression testing system integrated with deployment orchestration.
\alwayson{} and \activemigrate{} class of services deploying to the canary zone must pass \ufa{} failover regression tests where traffic is blocked to all services of \restorelater{}  and \terminate{} class for five minutes.
The system analyzes application metrics during isolation to detect new fail-close dependencies, automatically rolling back a deployment if it compromises \ufa{} safety guarantees.

During a 45-day trailing observation period, the canary regression gate analyzed approximately 8,000 deployments per week and identified three regressions.
The sparsity of defects found during canary testing is a testament to the power of static analysis, which finds defects sooner in the CI pipeline.

If any of the checks/tools find a fail-close violation between a higher-priority service and a lower-priority \restorelater{} or \terminate{} service, then the lower-priority service is off-boarded from \ufa{} termination until the team for the higher-priority service resolves the violation.

\textbf{AI-Powered End-to-End Validation.} The final layer integrates a GenAI-powered mobile testing platform~\cite{uber2024dragoncrawl,marcano2026scalingmobilechaostesting} with a fault injection system to provide comprehensive end-to-end validation of \ufa{}'s impact on critical use-cases.
This combination addresses the challenge of validating \ufa{}'s fail-open assumptions across complete customer experiences without requiring manual test authoring or risking production stability.
The testing platform autonomously navigates mobile applications using high-level intent specifications, while the fault injection system simulates downstream failures where calls to \restorelater{} and \terminate{} class of services are blocked.

The integrated system operates through nightly regression testing and pre-drill validation, achieving 99.27\% pass rates compared to 98.39\% for traditional tests while requiring no maintenance.
When tests fail under fault injection, the system correlates failures with specific service pairs and opens actionable tickets for service owners to address dependency violations.
This automation ensures that \ufa{} dependency violations are systematically identified and resolved.

The end-to-end validation system onboarded 47 core flows, which represent the most critical customer experiences to run with fault injection for \ufa{}.
The combined system identified 23 resilience risks that would have impacted customers and conducted pre-validation for 43 UFA drills.

\section{Phased Rollout}
\label{sec:rollout}

Deploying \ufa{} across 6,000+ microservices and hundreds of engineering teams required a systematic, risk-managed approach spanning two years.
The rollout followed a carefully orchestrated progression from \terminate{}-class non-production workloads to increasingly higher-priority service tiers, with each phase gated by readiness reviews and drill validation.
Phase 1 (2024) targeted non-production environments and T3-T5 services, reducing capacity from 2.0$\times$ to 1.65$\times$ while proving the model's viability with minimal business risk. Phase 2 (2025) expanded to T2 services and will include selective T1 oversubscription, reaching the ultimate goal of 1.3$\times$ capacity.

The phased approach proved essential not only for risk management but also for tool maturation and integration. Early phases relied primarily on runtime dependency analysis and basic traffic isolation, while later phases incorporated the full suite of safety tools, including static analysis and AI-powered end-to-end validation.
Each tier progression required extensive drills to validate both the architecture and our processes, with lessons from earlier phases informing the design of subsequent rollout strategies.

\section{Quantitative Impact of \ufa{}}
\label{sec:impact}

\begin{figure}[!t]
    \centering
    \includegraphics[width=\linewidth]{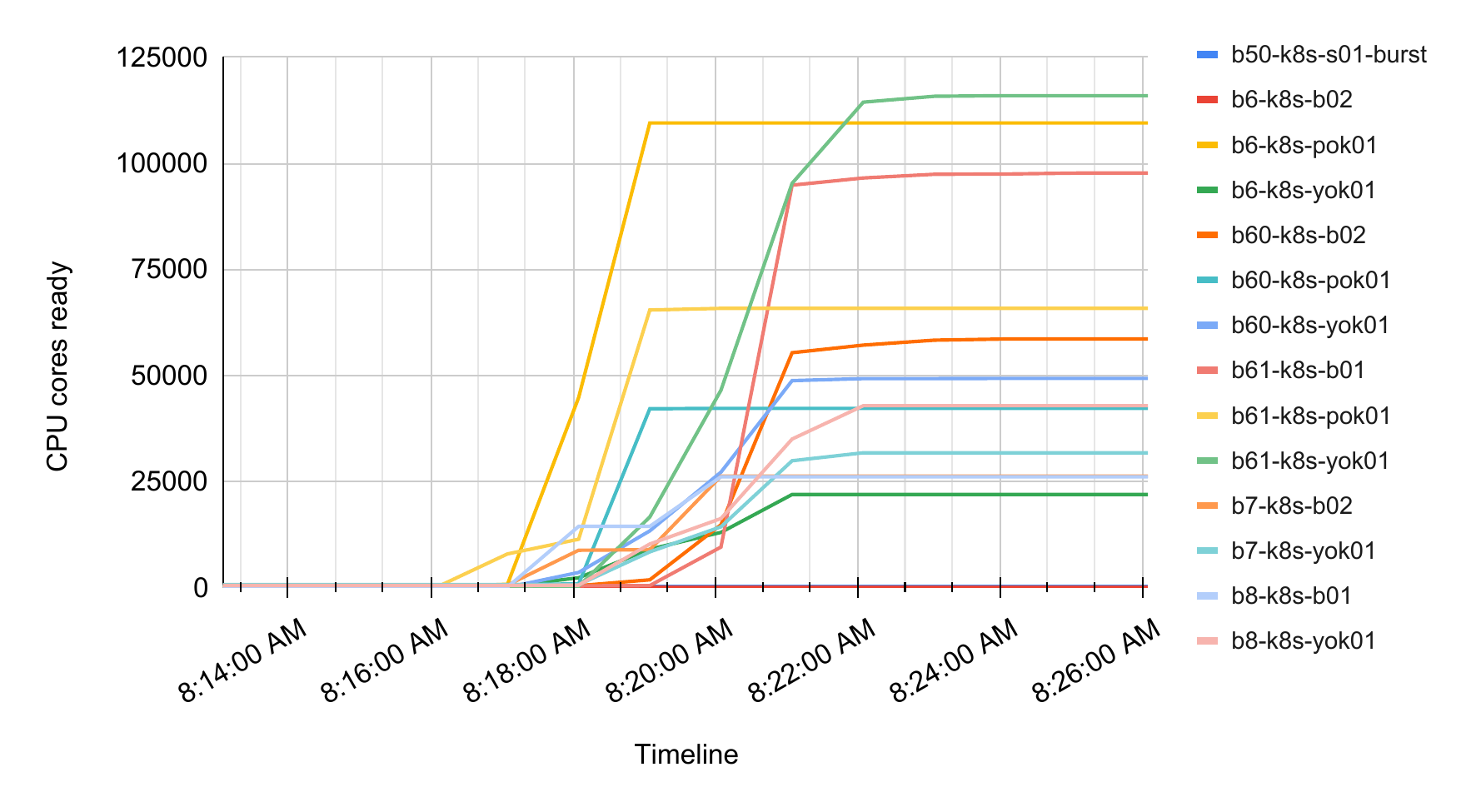}
    \caption{Demonstration of capacity conversion from batch to failover during a real 17-hour failover event. 
    Different colored lines represent different burst clusters.}
    \label{fig:bustcapacity}
\end{figure}

\begin{figure}[!t]
    \centering   \includegraphics[width=\linewidth]{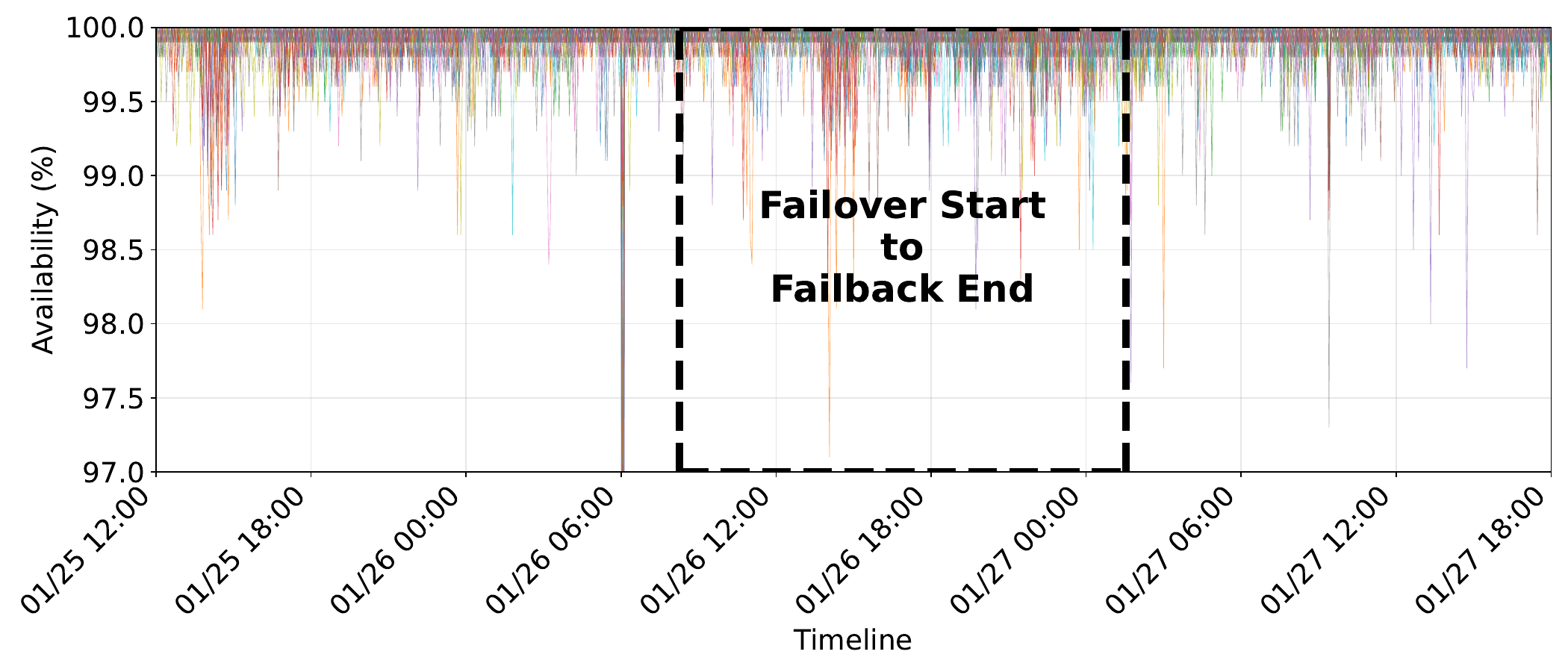}
    \caption{The overall average availability of core trips for the top 100 cities remains consistently above 99.97\%. The highlighted failover window (1/26/2026 8:15 AM - 1/27/2026 1:30 AM) shows no degradation in city-level availability compared to the period outside the failover.}
    \label{fig:availability2}
\end{figure}

We now present empirical evidence from production deployments and a real-world failover situation that validates the system's core design.
We choose to show a failover that happened on Jan/26/2026 at 8:15 AM lasting for around 17 hours.

\textbf{Rapid Capacity Conversion.}
Figure~\ref{fig:bustcapacity} demonstrates \ufa{}'s ability to rapidly convert batch capacity to burst capacity during the 17-hour failover.
For the failover initiated at 8:15 AM, the full capacity was available in about eight minutes  by 8:22 AM.
The different-colored lines represent various availability zones within the failover region, indicating consistent performance.
This rapid resource availability is critical for \ufa{} to achieve its SLAs.

\textbf{Service Availability During Failover.}
The overall availability of our services across the top 100 cities is shown in Figure~\ref{fig:availability2}, along with the time window during which the failover and failbacks occurred. For reference, the average availability before the failover is 99.97\% and remains at 99.97\% throughout the failover and failback windows.
Furthermore, the availability remains at 99.97\% during the first ten minutes of the failover, during which the capacity conversion happens, and also during the first thirty minutes when not all \restorelater{} instances are ready yet.
This demonstrates that \ufa{}'s differentiated SLA model maintains service quality for critical workloads even during the most resource-constrained periods of a failover.

\begin{figure}[!t]
    \centering   \includegraphics[width=\linewidth]{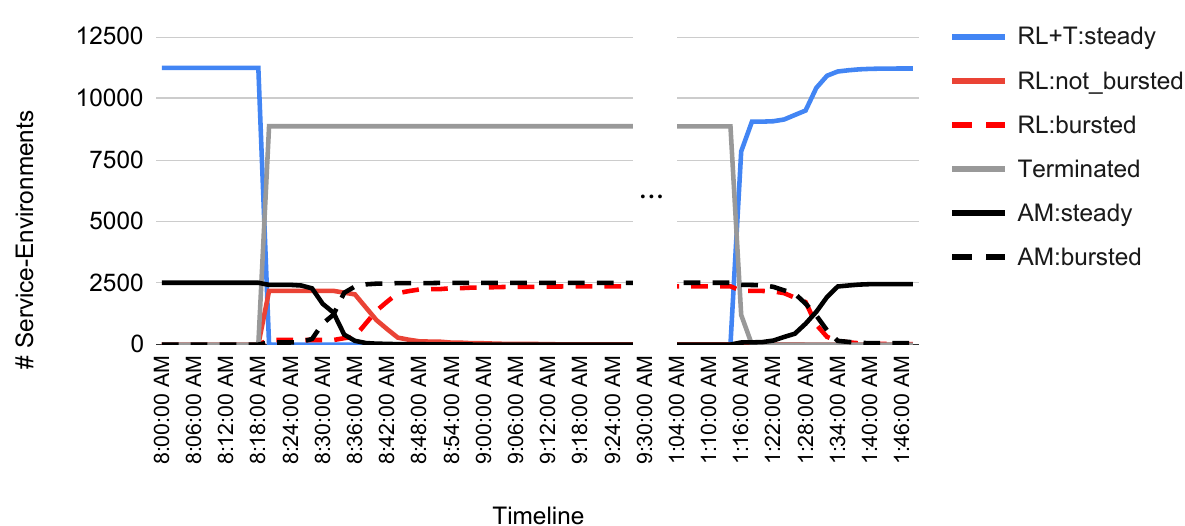}
    \caption{Conversion of various containers during a real failover from 8:20 AM to failback completion at next day 1:30 AM (17 hours).}
    \label{fig:failoverconversion}
\end{figure}

\textbf{Container Orchestration at Scale.}
Figure~\ref{fig:failoverconversion} illustrates the conversion of containers of various failure classes during the failover starting at 8:15 AM. Failback starts the next day at 1:15 AM and completes at 1:30 AM.
It highlights the coordinated movement of services with different reliability profiles during the failover process, with \alwayson{} and \activemigrate{} services maintaining priority access to resources.

The solid blue line (\texttt{RL+T:steady}) represents the
\restorelater{} and \terminate{} class
service environments (SEs) running under a steady state (before failover).
It drops sharply right after the failover begins at 8:15 AM, since they are terminated immediately to free resources.
It stays at zero until failback restores it the next day, starting at 1:15 AM and finishing at 1:30AM.
The gray-colored \texttt{Terminate} represents the \terminate{} class of SEs that are never restored anywhere during the failover.

The solid black \texttt{AM:steady} line represents the \activemigrate{} class of SEs that are restored with priority and cannot be terminated without restoring in the burst capacity.
It starts to drop in synchrony with its complement dotted black \texttt{AM:bursted} line.
As new \texttt{AM:bursted} containers are brought up, the old ones in \texttt{AM:steady} are terminated.
This conversion finishes within the first 15 minutes (8:30 AM).
The converse is true the next day, 1:15 AM, when the failback begins.

The solid red-line \texttt{RL:not\_bursted}  and the dotted red-line \texttt{RL:bursted} are complementary lines belonging to the \restorelater{} class.
\texttt{RL:not\_bursted} spikes as soon as the \restorelater{} containers are terminated and stays at that level for about 15 minutes (until 8:25 AM) until the burst capacity comes online. At that point, it starts to drop and its complement, \texttt{RL:bursted}, starts to rise.
\texttt{RL:bursted} stays for the duration of the failover and starts to decrease as the failback begins.
Unlike \activemigrate{}, these don’t drop immediately.
They maintain a stable baseline until new capacity is activated.

\begin{figure}[!t]
    \centering   \includegraphics[width=\linewidth]{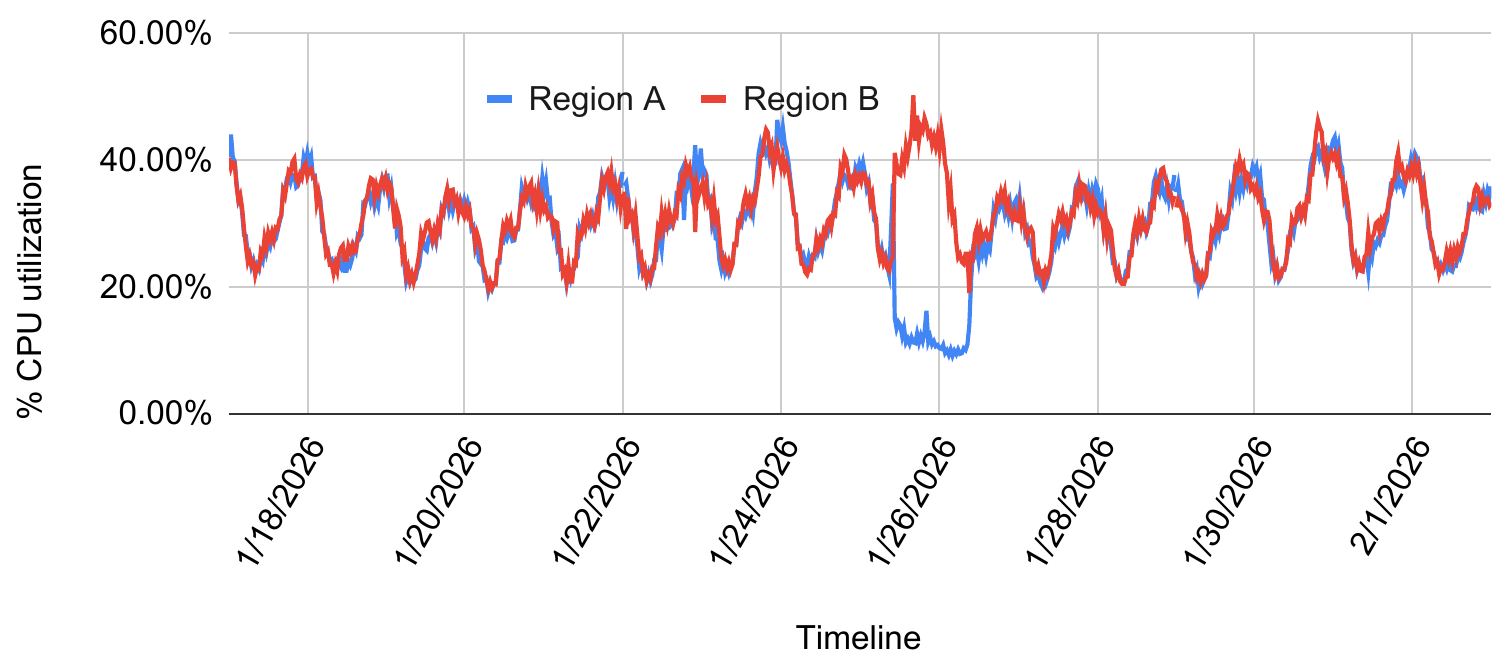}
    \caption{CPU utilization of two regions over time, showing the peaks during a 17-hour failover on 1/26/2026. RegionB sustains a peak failover reaching an average utilization of 50.2\%.}
    \label{fig:peakcpu}
\end{figure}

\paragraph{CPU Utilization During a Failover.}
Figure~\ref{fig:peakcpu}  illustrates regional CPU utilization during the failover.
RegionB sustained a peak failover on Jan/26/2026 with an average utilization reaching 50.2\%. Importantly, utilization remained within safe operational thresholds, demonstrating that the \ufa{} design enabled effective capacity utilization without overshooting high-risk levels. The result confirms that failover buffers can absorb peak loads while preserving headroom and stability for critical services.

 To evaluate system stability during UFA failover, we further analyzed the frequency of container evictions triggered by host CPU utilization exceeding the 75\% safety threshold, comparing these results against a non-UFA baseline. 
 Within a deployment of 850K pods, the peak eviction rate during failover reached 312 per hour—approximately $2\times$ the baseline peak of 160 per hour. 
 This spike was heavily concentrated within the initial hour of the failover event; for the remainder of the period, eviction rates subsided to near-zero, aligning with the overall average observed during standard operating conditions. 
 While the peak eviction count was elevated, it represented only a negligible fraction of the total deployment, ensuring that overall service availability remained unimpacted throughout the transition.

\begin{figure}[!t]
\begin{minipage}{0.45\textwidth}
    \centering   \includegraphics[width=\linewidth]{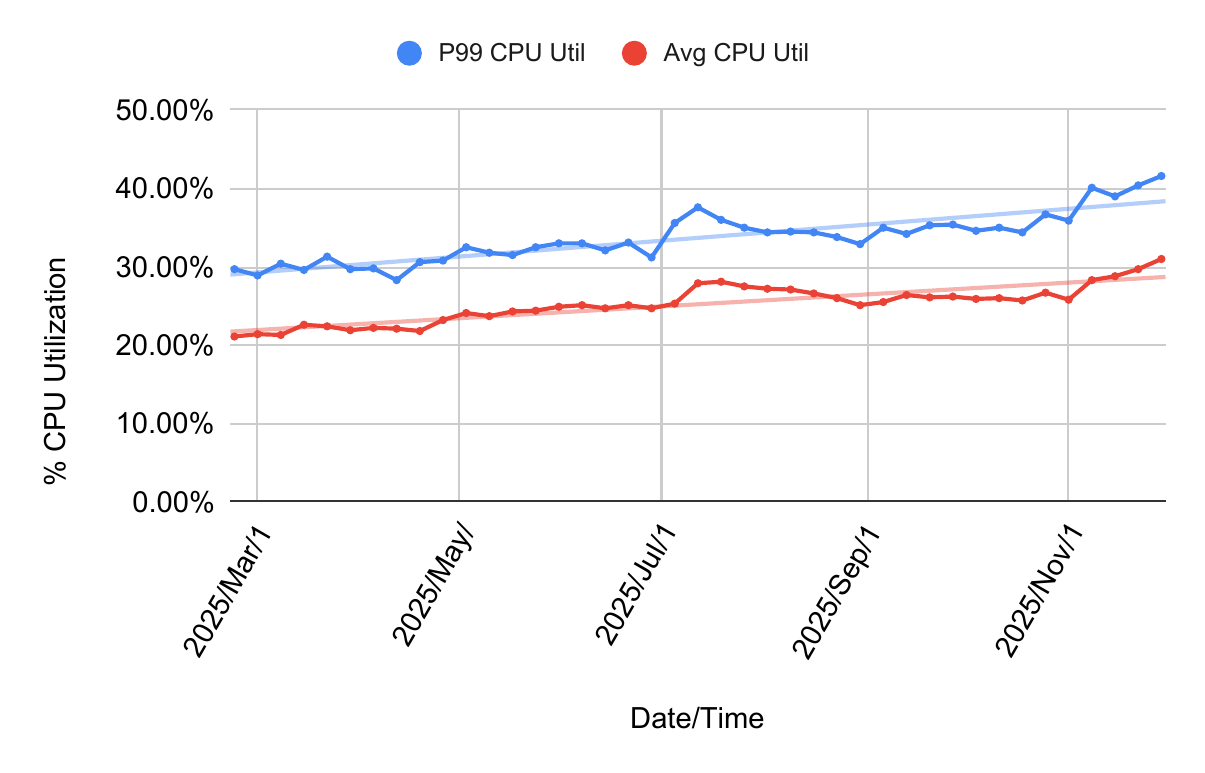}
    \caption{Global CPU utilization grew from an average of $20\%$ (P99 $30\%$) to $31\%$ (P99 $42\%$).}
    \label{fig:cpuutilization}
\end{minipage}
\end{figure}

\textbf{Fleet Utilization Improvements.}
Figure~\ref{fig:cpuutilization} demonstrates that global CPU utilization increased from an average of 20\% (P99 30\%) to 31\% (P99 42\%), representing a noticeable improvement in resource utilization during the 11 months as UFA was rolled out to the fleet.
This increase in utilization directly translated to cost reduction.
During this time, 1.025M cores were reduced through \ufa{}'s phased resource release.

A dissection of the 1.025M cores saved revealed that $\sim$550K (54\%) cores were attributed to the break-before-make strategy and another $\sim$475K (46\%) cores to the make-before-break strategy, signifying the importance of both approaches.

\section{Experience and Lessons Learned}
\label{sec:lessons}

\ufa{}'s deployment across 6,000+ microservices and hundreds of engineering teams revealed operational insights that extend beyond the technical architecture.
This section distills key lessons learned from over a year of production deployment, systematic drills, and systems development.

\textbf{Global Scale Brings Unexpected Complexity.} Despite comprehensive testing, Uber's global diversity surfaced challenges impossible to anticipate in staging.
The platform behaves differently from country to country and city to city due to varying regulations, cultural patterns, and operational constraints.
During drills, we discovered location-specific issues, such as longer driver response times at certain airports, that our broad testing missed. This reinforced that no amount of pre-deployment validation can substitute for careful, graduated production rollout with comprehensive monitoring across all operational contexts.

\textbf{Fast Regional Failover Mitigates Gray Failures.}
In practice, \ufa{} benefits from Uber's existing practice of proactively initiating regional failovers as soon as gray failures~\cite{GrayFailures2017} are detected, rather than waiting for a full outage.
Once a failover begins and traffic shifts to the target region, \ufa{} is triggered and operates normally.
We found that even when the target region experiences minor degradation, the impact is contained to non-critical services draining to burst clusters and does not cascade to critical services.
We also learned that when failures originate from \ufa{}-managed services themselves, restarting them in burst clusters often resolves the issue, since the services come up on fresh infrastructure that eliminates non-code-related problems.

\textbf{The Critical Role of Shift-Left Safety Tooling.} The multi-layered safety approach proved indispensable, but with some nuance.
Our static analysis capabilities have been key to preventing regression by enabling developers to quickly identify fail-close dependencies during development.
By shifting left, we not only achieve high reliability initially but also maintain it long-term as teams continue deploying changes.
However, AI-powered end-to-end testing revealed both opportunities and limitations.
While it simplifies mobile testing by enabling specification through natural language intents, its LLM-driven action selection can introduce flakiness, complicating fault attribution.
We learned to establish stable baseline tests without fault injection, then use differential analysis against fault-injected executions to isolate failures attributable to \ufa{}.

\textbf{The Staging-Production Gap.} Testing failover workflows across multiple infrastructure layers is inherently risky.
An isolated staging environment was essential as the foundation for automated tests, but the gap between staging and production complexity remained significant.
Systematic production drills and comprehensive observability were critical to bridging this gap.

\textbf{Failure Class Hygiene Improves Quality.} Each failure class carries different expectations for rollout safety, testing, and monitoring. \ufa{} prompted service owners to re-evaluate and re-classify their services, which in turn raised operational standards to match the rigor expected of each class.

\textbf{Engineering Investment and Adoption.}
Rolling out \ufa{} required substantial engineering and organizational effort.
Principal investigators had to secure buy-in from all organizations, addressing nervousness around failure and positioning \ufa{} as an Uber-wide reliability program.
A centralized platform team of about 50 engineers drove the initiative, but adoption required every product organization and team to remediate fail-close dependencies through the readiness review process.

Three main strategies emerged for addressing fail-close dependencies:
(1) code-level fixes to convert fail-close dependencies into fail-open,
(2) up-tiering a callee service with direct involvement from the \ufa{} program team, and
(3) service re-design.
For example, Uber's delivery business introduced ``Lite Mode'' feeds to ensure continuity during failovers, which is similar to Defcon's approach at Meta~\cite{DefconOSDI23}. This saves tens of thousands of CPU cores.
About 200 previously lower-priority services were re-tiered into the \activemigrate{} class.

\textbf{Limits to on-demand cloud capacity.}
Cloud providers cannot guarantee on-demand capacity at the scale \ufa{} requires (\S~\ref{sec:cloud}). Engaging with providers earlier to contractually secure some burst capacity could help. However, an inherent tension exists: cloud providers also prefer to maximize utilization and are reluctant to reserve large idle capacity at affordable rates. Fully relying on cloud burst buffers for failover is therefore not realistic.

\textbf{Retrospective: What We Would Change.}
With the benefit of hindsight, two decisions stand out as areas we would approach differently.

\emph{(1) Avoiding batch displacement during failovers.}
\ufa{} currently evicts batch workloads to convert batch clusters into burst capacity (\S~\ref{sec:batch}). An alternative is to coordinate with service owners to temporarily avoid scheduling non-critical services on batch clusters during the few hours of a regional failover. This would eliminate the batch capacity bottleneck without permanently reserving extra capacity.

\emph{(2) Cleaning up inter-service dependencies.}
The internal dependency graph between services is complex and not always well understood. Cleaning up these dependencies would yield a clearer structure and more predictable behavior during service termination. While this would ideally precede \ufa{}, it is a multi-year effort whose scope must be carefully weighed.

\textbf{Open Problems and Future Research Directions.}
\ufa{}'s deployment surfaced several challenges that we believe are relevant beyond Uber and warrant further research.

\emph{Certifying fail-open behavior at scale.}
Our three-layer pipeline (runtime analysis, static analysis, and canary regression testing) has proven effective, but scaling it to thousands of services required significant engineering effort. General-purpose tools for automatically certifying fail-open behavior across large service meshes remain an open problem.

\emph{Guaranteed elastic capacity at hyperscale.}
\ufa{} relies on pre-negotiated burst capacity because on-demand cloud provisioning cannot guarantee hundreds of thousands of cores within minutes. New infrastructure primitives for rapid, guaranteed elastic capacity at hyperscale would reduce the need for such pre-negotiation. This would also improve steady-state utilization: with more guaranteed burst capacity available during failovers, more services can be safely placed in overcommit pools, raising fleet-wide utilization beyond its current levels.

\emph{Sacrificing batch for real-time availability.}
\ufa{} repurposes batch clusters as burst capacity during failovers, temporarily displacing analytics and ML training workloads. The tradeoffs behind this pattern, including impact on batch SLAs and job restart costs, deserve further study. A related opportunity is selectively bursting only the services that truly need failover capacity, rather than displacing all batch workloads. This would lead to more balanced capacity usage during failovers and reduce unnecessary disruption to batch jobs.

\emph{Extending to disaggregated storage.}
As storage architectures move toward disaggregation, storage starts to resemble stateless services. This opens the possibility of applying \ufa{}'s oversubscription and differentiated SLA principles to the large storage fleet, which is currently outside \ufa{}'s scope.

\section{Related Work}
\label{sec:related}

\ufa{} builds on research in distributed systems reliability, failure handling, and resource management.

\textbf{Data-Driven Reliability Analysis.} Like \ufa{}, several studies have leveraged operational data to inform design decisions. Microsoft Azure built an online oversubscription prediction engine using workload characteristics~\cite{ResourceCentral2017}; our over-commit simulator is similar in intent, though it operates offline.
Google's storage system availability analysis~\cite{osdi2010} used production data to evaluate design choices, similar to how we analyzed four years of failover data to justify differentiated SLAs. Hardware failure studies~\cite{dram2009,disks2007} demonstrate the value of empirical analysis in understanding real-world failure patterns. The ``Designing for disasters'' work~\cite{disasters2004} shares our philosophy of using financial objectives to automate disaster-tolerant system design.

\textbf{Failure Detection and Recovery.} \ufa{}'s reliance on human operators contrasts with automated systems like FALCON~\cite{falcon2011} and Pigeon~\cite{nsdi2013}. We deliberately chose human-in-the-loop failover due to the complexity and cost implications of regional failovers~\cite{GrayFailures2017}, aligning with network reliability studies~\cite{sigcomm2011,prefix2018} that acknowledge operator roles in large-scale infrastructure.

\textbf{Multi-Tenant Resource Management.} Co-locating different service tiers or batch with stateless has been extensively explored~\cite{Mesos2011,Quasar2014, Heracles2015, retro2015, borg2015, zhang2016history,BorgOmegaK8CACM2016, AutothrottleNSDI2024}.
The emphasis of these prior works is on performance isolation.
However, \ufa{}'s differentiated reliability model requires services to maintain fail-open behavior when lower-priority services are preempted, a dependency safety requirement beyond traditional resource fairness.

Meta's Defcon~\cite{DefconOSDI23} takes a complementary approach and performs graceful feature degradation \emph{within} services during overload to protect availability. \ufa{} instead preempts entire non-critical service containers and combines workload-level shedding with cross-tier dependency safety. Phoenix~\cite{PhoenixASPLOS2025} similarly sheds non-critical containers using criticality tags, but does not address cross-tier dependency isolation.

Oversubscribing limited resources, whether execution units~\cite{tullsen1998simultaneous},  memory~\cite{ESXWaldspurger}, power~\cite{PowerFan}, network~\cite{greenberg2009vl2, al2010hedera}, to name a few, is a popular method to increase utilization and reduce idleness.
\ufa{} follows a similar idea, but the tiered SLA and fail-close resiliency are unique.

\textbf{High Availability Architectures.} Traditional approaches like Remus~\cite{remus2008} provide transparent VM-level failover but treat all software uniformly. Google's CAPA~\cite{capa2024} shares our emphasis on dependency safety through its ``Low-Dependency Requirement.'' \ufa{}'s microservice-granular approach enables more flexible failover strategies and differentiated SLAs that optimize for both reliability and efficiency.

\textbf{Correlated Failures and System Resilience.} Research on correlated failures~\cite{correlated2020, independenceAsService2014, nsdi2006} highlights challenges that \ufa{} addresses through its comprehensive drill programs and numerous tools (Section~\ref{sec:tools}). Our approach considers service-level dependencies and their impact on cascading failures. 
Tang et al.~\cite{FailThroughCracks2023} show that such cross-system interaction failures are a leading cause of production incidents at Google, Azure, and AWS.
The Alibaba Cloud work on rapid low-cost failover~\cite{failover2017} shares similar concerns about massive-scale reliability but emphasizes state recovery.

\textbf{Microarchitectural Optimizations for Microservices.}
Prior work has examined microservice inefficiencies and proposed optimizations at both the system and microarchitectural levels~\cite{song:isca22:thermometer, khan:micro21:twig, ripple-khan21, softsku, duplexity, asriraman18, accelerometer, bubbleupMars}. For instance, SoftSKU~\cite{softsku} tunes server configurations to microservice needs, while Accelerometer~\cite{accelerometer} identifies acceleration opportunities for common orchestration overheads.
 Orthogonal to these approaches, \ufa{} focuses on differentiated SLAs and CPU oversubscription with failure isolation to improve utilization.

\section{Conclusions and Future Work}
\label{sec:conclusion}

\ufa{} demonstrates that the long-standing tension between reliability and efficiency in hyperscale systems is not inevitable.
Guided by a data-driven insight that catastrophic failovers occur less than 20 hours annually, we transformed a wasteful 2$\times$ capacity model into an intelligent oversubscription architecture that has already returned over a million CPU cores.
The true complexity was not in any individual technical innovation, but in orchestrating a socio-technical transformation across 6,000 microservices developed by hundreds of independent teams through systematic drills, rigorous validation, and multi-layered safety tooling.

Future \ufa{} extensions will go beyond stateless services to offer differentiated SLAs for stateful services.
Several open directions remain: combining static analysis with generative AI to automatically fix fail-close issues,
developing general-purpose tools for certifying fail-open behavior at scale,
and working with cloud providers toward guaranteed elastic capacity at hyperscale.
We hope our work inspires further research in this space.

\section*{Acknowledgments}

We thank people from various teams at Uber for their invaluable contributions to this project: 
Abhishek Jha,
Aditya Jain,
Arturo Bravo Rovirosa,
Arun Krishnan,
Christoffer Hansen,
Darshil Kapadia,
Deepanker Sachdeva,
Egor Grishechko,
Eric Chin,
Gaurav Bansal,
Haarith Devarajan,
Jeffrey Chang,
Kamran Massoudi,
Kamran Zargahi,
Krishna Gupta,
Lingchao Chen,
Long Zhen,
Matt Loughney,
Milos Radic,
Monojit Dey,
Pawandeep Singh,
Prasanna Vijayan,
Ren\'e Just,
Riikka Lahenius,
Sarah Ahmed,
Scott Brown,
Sonal Mahajan,
Srikar Paruchuru,
Sumit Singh,
Ting Wang,
Vlad Sandu,
Yufan Xu, and
Yuxin Wang.
We also thank David A. Maltz, our paper shepherd, for his feedback and support.

\newpage

\bibliographystyle{plain}
\bibliography{ref}

\end{document}